# A Solar cycle correlation of coronal element abundances in Sun-as-a-star observations

David H. Brooks[1,*], Deborah Baker[2], Lidia van Driel-Gesztelyi[2,3,4] & Harry P. Warren[5]

[1]College of Science, George Mason University, 4400 University Drive, Fairfax, Virginia 22030, USA

[2]Mullard Space Science Laboratory, University College London, Holmbury St. Mary, Dorking, Surrey, RH5 6NT, UK

[3]LESIA-Observatoire de Paris, CNRS, UPMC Univ. Paris 06, Univ. Paris-Diderot, Meudon Cedex, Paris, F-92195, France

[4]Konkoly Observatory of the Hungarian Academy of Sciences, H-1121 Budapest, Hungary

[5]Space Science Division, Naval Research Laboratory, 4555 Overlook Avenue SW, Washington, District Of Columbia 20375, USA

*Correspondence to: dhbrooks@ssd5.nrl.navy.mil
 Present address: Hinode Team, ISAS/JAXA, 3-1-1 Yoshinodai, Chuo-ku, Sagamihara, Kanagawa 252-5210, Japan

**Abstract:**

The elemental composition in the coronae of low-activity solar-like stars appears to be related to fundamental stellar properties such as rotation, surface gravity, and spectral type. Here we use full-Sun observations from the Solar Dynamics Observatory, to show that when the Sun is observed as a star, the variation of coronal composition is highly correlated with a proxy for solar activity, the F10.7 cm radio flux, and therefore with the solar cycle phase. Similar cyclic variations should therefore be detectable spectroscopically in X-ray observations of solar analogs. The plasma composition in full-disk observations of the Sun is related to the evolution of coronal magnetic field activity. Our observations therefore introduce an uncertainty into the nature of any relationship between coronal composition and fixed stellar properties. The results highlight the importance of systematic full-cycle observations for understanding the elemental composition of solar-like stellar coronae.

**Introduction**

Knowledge of the elemental composition of the Sun underpins our understanding of the transport of energy from the deep interior, through the outer atmosphere, and into the heliosphere. The mass supplied to the solar wind and the Sun's outer atmosphere originates in the lower atmospheric layers, yet surprisingly, the elemental composition in the corona and slow solar wind is different than in the photosphere *(1)*: low first ionization potential (FIP) elements that are easy to ionize in the chromosphere are preferentially enhanced (fractionated) by factors of 3–4 *(ref 2)*, whereas high-FIP elements that remain neutral in the chromosphere retain their photospheric abundances. Furthermore, the composition appears to vary from atmospheric structure to structure, making it a potential signature of the coronal heating process. This composition enhancement process is known as the FIP effect *(3)*, and it is frequently used as a powerful diagnostic in solar and stellar astrophysical studies *(4)*.

Recent observations have provided evidence of cyclic variability of elemental abundances in the solar wind *(5,6,7)*. Observations from the Wind spacecraft show that the He abundance in the slow wind is strongly correlated with sunspot number *(8)*, and decreases at solar minimum in the fast wind *(6)*. In situ measurements made by the Solar Wind Ion Composition Spectrometer instrument on the Advanced Composition Explorer observatory confirm that some composition measures vary with the 11-year solar activity cycle. For example, the abundance ratio of the high-FIP elements Ne and O was enhanced by about 40% in the slowest solar wind streams (<400 km s$^{-1}$) during the last solar minimum from 2007 to 2010 compared to during the previous solar maximum between 2000 and 2003 *(ref 9)*. The formation of the slow solar wind is not yet understood, so these observations alone cannot tell us whether this cyclic variation is due to properties of the wind, once fully heated and accelerated, or to cyclic properties detectable in the wind source regions. More recent measurements made remotely using the Solar Ultraviolet Measurements of Emitted Radiation instrument on the Solar and Heliospheric Observatory satellite, however, showed that a similar cyclic variation is detectable in the Ne/O ratio in coronal streamers *(10)*, a slow solar wind source candidate. This is highly suggestive that the variation of the slow wind with the solar cycle is a property of the formation site, though this evidence is also indirect as a link with the solar wind is inferred, and other coronal features that may be potential sources were not sampled. Importantly, if variations from photospheric to coronal composition with the solar cycle are detectable in full disk integrated observations of the Sun's atmosphere spectroscopically, rather than only by in situ measurements in the solar wind or in detailed features, then similar variations might be detectable spectroscopically on solar-like stars.

Here we show that, when the Sun is observed as a star, the coronal composition is highly correlated with the F10.7 cm radio flux during the solar cycle between 2010 and 2014. We also discuss the implications of this result for observations of cyclic coronal composition variations in solar-like stars.

**Results: Solar Observations**

Whether the implications for stellar observations are significant or not depends on bridging the solar–stellar connection more directly. We cannot make in situ particle measurements in any solar analog astro-sphere, and coronal streamers are detailed features that are not spatially resolved in stellar spectra. We can, however, determine whether a solar-cycle variation in coronal composition is detectable in full-disk integrated Sun-as-a-star spectra, using the data obtained by the Solar Dynamics Observatory (SDO) Extreme-Ultraviolet Variability Experiment *(11)* (EVE). These data also implicitly include all potential solar wind sources in the low atmosphere, making a direct link with the solar cycle easier since we make no assumptions about what the source is. We used EVE irradiance measurements to compute the ratio of coronal to photospheric composition, or FIP bias ($A_c/A_p$), for daily averaged spectra covering the period between April 2010 and May 2014 when EVE was fully functional (see "Methods" section for details). Note that we used irradiances from Fe, Mg, and Si lines to model the irradiance of the Ne VII 465.22 Å line. So the values, $A_c/A_p$, are measurements of the ratio of the coronal abundance derived from Fe, Mg, and Si ($A_c$), to the photospheric abundance of Ne ($A_p$). The latter is not directly measured in the Sun due to the absence of visible light spectral lines, and is therefore inferred from other sources *(2)*.

A good proxy for the 11-year solar cycle for comparison is the F10.7 cm radio flux *(12)*. In Fig. 1, we show the daily averaged solar radio flux for 1996–2016, with an orange overlay showing the period of our observations. The period covers from the end of the extended solar minimum of cycle 23, beyond the maximum of the current weak cycle 24. We also show a boxcar smoothed yearly average fit to the two cycles in red. The fit suggests that the F10.7 cm peak of cycle 23 was at least a factor of 1.4 higher than the F10.7 cm peak of cycle 24. The peak of cycle 24 was also at least a factor of 2.1 higher than the minimum of cycle 23.

Figure 2 shows the 27-day Carrington-rotation running average of the FIP bias measurements plotted against the 27-day running average F10.7 cm solar radio flux. We show these data to highlight any trends. The daily results are shown in Supplementary Fig. 1. The most significant new result, and a new source of information on the coronal heating process, is that the FIP bias derived from full-Sun spectra is clearly highly correlated with the F10.7 cm solar radio flux. The calculated linear Pearson correlation coefficient, r, is 0.88. There is a correlation not only in the general trend, but also in some of the small-scale details. See also Supplementary Fig. 2 for a direct correlation plot. The F10.7 cm radio flux is related to chromospheric activity, where the FIP fractionation is believed to take place, and this correlation also provides supporting evidence for that view. The ratio of coronal to photospheric composition increases from around 2.3 in mid 2010 to close to 4 in 2014; with the minimum being consistent with a relatively high (above photospheric) basal level for the quiet Sun FIP bias. On the basis of the strong correlation with the radio flux, and considering the uncertainties in the FIP bias measurements (see "Methods" section), this corresponds to a detection of a variation during the cycle at a level of $5.7\sigma$.

**Discussion**

Our finding of a coronal composition variation in Sun-as-a-star full disk-integrated spectra, which is correlated with the solar cycle, is intuitively consistent with the area coverage of different features (quiet Sun, active region, coronal holes), and their consequent contribution to coronal composition, changing with the solar cycle. In particular, the area coverage of active regions increases with activity.

Our observations also imply that the compositional enhancements are detectable in Sun-as-a-star spectra at temperatures as low as 0.5 MK. Previous studies have found the enhancement to be most visible above 1 MK *(13)*, though this may depend on the relative contributions of quiescent and active region plasma at different cycle phases. Our results are also consistent with previous detections at these temperatures in coronal streamers *(10)*.

When we observe the Sun as a star, we detect a co-variation with the solar cycle, which clearly shows that coronal composition varies with the evolution of coronal magnetic field activity; as found also in studies of solar active regions *(14)*. This finding has potentially important implications for stellar astronomy. Although the solar results should not be directly extrapolated to other stars, our observations imply that some solar-like stars may also have cyclic dependencies. The magnitude of the coronal abundance variation due to activity will likely depend on the cyclic properties of the star, and may also be different for different cycles; as suggested by the close correlation with the solar F10.7 cm radio flux we found for the Sun. Our results, however, support the view that stellar coronal abundances are influenced by coronal magnetic field activity and the heating processes, and are not simply determined by the properties of the star. Cyclic effects can be present, and care should be taken when interpreting observed coronal abundances and trends between different stars.

One recognized example of this, that we revisit here, is the apparent observed trend between coronal composition and spectral type *(15)*. Following conventions used by the stellar community, in Fig. 3, we plot the ratio of photospheric to coronal abundance logarithmically (log (Ap/Ac)) as a function of spectral type for a sample of stars *(15, 16)*. We show the Sun as a red dot. Note that, whereas our results are based on Ne alone, the stellar coronal values are averages from the high-FIP elements C, N, O, and Ne. Since the techniques used to derive the values are also somewhat different, the absolute values may not be directly comparable. The value of −0.6, corresponding to a FIP bias of 4, was measured close to solar maximum in 1969 *(ref 13)* and is similar to our measured maximum for solar cycle 24. There is a clear linear trend from solar-like enhanced coronal abundance in G-type stars to an inverse FIP-effect ( > 0.0) in the later spectral types. In the right panel, we show the range of solar abundance ratio values on the plot due to the solar cycle. The solid red line connects the value measured at solar maximum (−0.6) to the value measured at solar minimum (−0.37). Our measurements show a significant degree of variation that is large enough to be detectable on the Sun, introducing uncertainty, and complicating the interpretation of any relationship between coronal composition and basic stellar properties.

There have been several studies of the relationship between coronal composition and activity *(17, 18)*, but these generally use proxies for activity, such as the ratio of X-ray to bolometric luminosity, which are measured at a single observing time, and even X-ray luminosity can show a cyclic variation *(19)*. For example, in the case of the Sun, the X-ray luminosity varies from $2.7 \times 10^{26}$ erg s$^{-1}$ near solar minimum to $4.7 \times 10^{27}$ erg s$^{-1}$ near solar maximum; *(20)* more than an order of magnitude. If the solar variation in coronal composition, shown by the vertical bar in Fig. 3, is representative of the activity cycles of the F5–K5 stars, then essentially our results imply that the positions of at least these stars in the plot dynamically change with the cycle. As they would also do in a coronal composition–activity plot (if the cyclic evolution of X-ray luminosity were used as the activity indicator), or in plots against combined activity and other stellar parameters such as previous authors have shown *(4)*.

We have attempted to associate the abundance ratio values in the plot with the phases of the stellar cycles of the corresponding object, but this proved difficult. Most of the long-term stellar chromospheric activity monitoring, e.g., from Mount Wilson *(21)*, ended prior to the launch of XMM-Newton and Chandra, making attempts to extrapolate to the observation dates insecure, due to the verified existence of multiple and modulated stellar cycles *(22, 23)*.

Furthermore, very active stars that show large inverse-FIP effects, such as AB Dor, EK Dra, and AU Mic, some of which do not follow the apparent trend, have been excluded from similar plots in the past because of difficulties in understanding the physics of fast rotators and the effects of close binaries *(16)*. We have re-introduced these stars to our plot (green squares) to show how their deviation from the trend compares to the magnitude of a solar-like cycle variation. While the FIP bias in EK Dra could be consistent with the scatter observed in early G spectral types, the AB Dor and AU Mic FIP biases appear to be smaller than the scatter in their respective K and M types.

We note that attempts to detect time-dependent compositional variations in a few stellar case studies have not yet been successful. AB Dor, for example, was surveyed over a 15-month period by

XMM-Newton and Chandra, and no variation was found *(24)*. Our results suggest, however, that cyclic variations may only be observed on longer time-scales. Figure 2 implies that the Sun would show approximately the same coronal value between 2012 and 2014, with a significant shift only over a longer time-scale, perhaps half a cycle; and the chromospheric cycle for AB Dor is 17 years *(25)*.

Apart from the long-term cyclic trends, there is clear variability in coronal composition on shorter timescales of days. Some of these changes may be due to flaring activity, but similar effects due to the passage of active regions have also been seen in stars, e.g., EI Eri *(26)*. To truly understand stellar coronal abundances, the composition enhancement process, and the heating of solar and stellar coronae, an observing strategy of re-observation and regular monitoring of selected stellar targets, including those with known activity cycles, may be a fruitful avenue to pursue rather than the wide practice of collecting a large sample and comparing properties.

We suggest that sampling and comparison of consistent measurements are also important. The cluster of stars close to spectral type M5 include fully convective flare stars, and studies of these and other very active stars such as AR Lac *(27)* find that they do not show cyclic variations. These stars may still be a good sample for a separate comparison with fixed stellar properties, or even a general study of the relationship with activity level, albeit that the coronal compositions of this subset of stars may not change with activity or stellar properties. For the solar-like stars with well-defined activity cycles, however, comparison of observations made at the same cycle phase may uncover new information on the dependence on activity or stellar properties. This applies both to coronal composition measurements, and also proxies of activity such as X-ray luminosity.

Although based on monitoring of only part of the solar cycle (4 years), this is the first observation of a correlation between the variation of coronal abundances and the phase of the activity cycle in Sun-as-a-star observations. This strong correlation between FIP bias and radio flux is encouraging. If it continues to hold for lower and higher levels of activity, then the solar cycle variation may be larger still. Although outside the time-frame of our actual measurements, based on the ratio of the cycle 24 peak to minimum flux (see discussion of Fig. 1), the Sun may have been positioned around −0.27 during the cycle 23 minimum (corresponding to a FIP bias of 2). At the other extreme, since the current solar cycle 24 is relatively weak compared to cycle 23, the relative strengths of the two cycles (taken again from Fig. 1) suggest that the Sun may move as low as −0.73 (corresponding to a FIP bias of ~ 5). Once the detailed relationship between FIP bias and radio flux is understood from future work, the F10.7 cm flux may become a useful proxy for determining coronal abundances in the Sun-as-a-star. Previous work has shown that another proxy for solar activity (sunspot number) can be used to predict abundance ratios (Ne/O) in the solar wind *(9)*.

We hope that our observations will inspire novel ideas and new observing strategies to further our understanding of stellar coronae in the coming years.

## Methods

### Comparison of coronal composition and F10.7 cm radio flux

For completeness, Supplementary Fig. 1 shows the daily running averaged ratio of the coronal to photospheric composition. Note that we have not performed any special filtering for flaring periods. Since EVE observations of flares show that they evaporate photospheric plasma *(28)*, the ratio, Ac/Ap, may reduce during flares. Conversely, the F10.7 cm radio flux increases during flares, so the correlation between the two parameters may be lower i.e., opposite to the overall trend. We also show a direct correlation plot of the 27-day Carrington rotation running averaged Ac/Ap ratio and the F10.7 cm radio flux in Supplementary Fig. 2. The correlation is strong (r = 0.88) and shows potential evidence of at least a linear trend, though possibly also a rapid non-linear increase at lower values of the F10.7 cm flux, and a leveling off, or saturation, at higher values, which would be consistent with the cyclic variations we see in Fig. 2. Since our observations do not cover the complete solar cycle, we cannot determine the detailed relationship, so for the discussion in this initial study, we have adopted the simple approach of making rough estimates based on the relative strength of the F10.7 cm solar maxima and minima.

### Data sources

The F10.7 cm solar radio flux data were downloaded from the OMNI Space Physics Data Facility at NASA Goddard Space Flight Center. They are in units of $10^{-22}$ W m$^{-2}$ Hz$^{-1}$ and come originally from the National Oceanic and Atmospheric Administration's National Geophysics Data Center. The F10.7 cm data measure the total radio emission, within a 100 MHz band, from emission sources in the upper chromosphere and low corona distributed over the solar disk. The measurements are accurate to a few percent *(12)* and are obtainable from the ground in all weather conditions. This makes the F10.7 cm flux a very accurate indicator of solar activity.

SDO *(29)*/EVE has twin grating spectrographs that make measurements with 1 Å spectral resolution in the 50–370 Å and 350–1050 Å wavelength ranges covering a region of the EUV solar spectrum rich in spectral lines from O, Mg, Ne, Si, S, and Fe. The EVE data were downloaded from the website of the Science Processing Operations Center at the Laboratory for Atmospheric and Space Physics in Boulder, Colorado. The data are level-2 spectra merged from the two spectrographs and are fully calibrated 10 s integrated spectral irradiance measurements in units of W m$^{-2}$ nm$^{-1}$. The data are provided as hourly datasets containing 360 spectra. We averaged all of the hourly datasets within each 24 h period to produce daily averaged spectra, and converted the units to erg cm$^{-2}$ s$^{-1}$. We also assumed a 20% error on the irradiances due to the intensity calibration *(11)*, and performed our analysis on 1475 daily averaged spectra in total covering the April 2010 to May 2014 period.

### Spectral analysis

We extracted the irradiances of spectral lines covering a wide range of temperatures (Fe VIII–Fe XVI covering 0.5–2.8 MK) using software developed for analysis of EVE observations of flares *(28)*. This method integrates the observed intensity between two pre-defined wavelength limits. See Supplementary Figs. 3 and 4 for examples and the details of the lines used. These lines were selected because they are relatively clean and to attempt to minimize the impact of weak blends. We also show a much higher spectral resolution spectrum from the EUV Imaging Spectrometer *(30)* on Hinode for comparison. We used an EIS quiet Sun and an active-region line list *(31)* to assess the impact of blending on the EVE lines also detected by EIS. According to the lists, the blends within 1.5 Å of the line centroids contribute 0–25% of the intensity. This is comparable to the intensity error and the accuracy of the line-intensity reconstructions (see Supplementary Fig. 5 below). An exception is Fe XII 202.044 Å, where blends from surrounding Fe VII–XIII lines may contribute a slightly higher 30%.

For the critical Ne VII 465.221 Å line, used as the abundance diagnostic and shown in the bottom panel of Supplementary Fig. 4, there are blends near 464.3 Å and 466.2 Å (the Ca IX resonance line). In the case shown in the Figure, these two lines contribute <10% of the total intensity of the three profiles. The wavelength limits used for the extraction are set at 464.5 and 466.0 in this case, so much of the blended intensity is also excluded. By comparison with a triple Gaussian fit, we found that the integrated intensity is within 5% of the fitted intensity. Since Ca IX is formed at a slightly higher temperature than Ne VII, it is possible that the contribution of this line will increase at more active times of the solar cycle. So we performed a similar analysis for spectra taken in 2014. The contribution of Ca IX to the combined Ca IX plus Ne VII intensity does indeed increase from ~ 8 to ~ 13%, but the integrated intensity is still within 8% of the fitted intensity. At both quiet and active times, therefore, the difference is also well within the intensity error.

Note that most of the Ca IX contribution is actually excluded in reality because the significant part of the intensity lies outside the extraction limits. Also, these differences lead to an overestimate of the Ne VII intensity (albeit within the errors), which in turn would lead to an under-estimate of the magnitude of the FIP bias by a comparable percentage (5–8%) i.e., a cyclic variation from 1.5 to 4 would become a slightly larger cyclic variation from 1.6 to 4.3.

Note that the EIS spectrum, which is scaled and cleaned for display, is of a large patch of quiet Sun, so it does not include any contribution from active regions. The very hot Fe XVI 262.984 Å line is not detected, for example.

**Plasma composition measurement technique**

We adopted a method to compute the FIP bias that has been extensively tested on EIS data *(32, 33)* and modified it for EVE as follows. For each daily spectrum, we used the Si X 258.375 Å/261.04 Å density-sensitive ratio to compute the electron density, and used this density to calculate the theoretical spectral line contribution functions needed to derive the coronal temperature structure, or differential emission measure (DEM) distribution. We then determined the DEM using spectral lines from the low-FIP elements (Mg, Si, Fe) only and assuming photospheric abundances *(34)* (see example Supplementary Fig. 5). This method implicitly assumes a constant density relationship with temperature, which allows us to re-cast the full 2D temperature and density integral inversion as a 1D function of temperature alone. The assumption, however, introduces some uncertainty, since the method takes no true account of the density variation between the temperature regions where different lines are formed. Fortunately, most of the spectral lines we use are formed close in temperature to the Si X lines used to measure the density (within 0.1 dex). For spectral lines formed at higher temperatures, such as those of Fe XIV—Fe XVI, we expect the density to be lower, and we have verified that the total contribution function remains within the intensity error in this case. For spectral lines formed at lower temperatures, such as those of Ne VII, Si VII, Mg VII, and Fe VIII—Fe IX, the total contribution function remains within the intensity error until the density reaches an order of magnitude higher than we have measured in the complete data set. Note that the critical Ne VII 465 Å line, and the Fe VIII 131 Å line that principally constrains it, remain within the intensity error for all densities.

Numerically, the DEM was computed using a Markov–Chain Monte Carlo (MCMC) algorithm *(35, 36)*. We then used the derived DEM to compute the irradiance of the clean and unblended Ne VII 465.221 Å line, which comes from the high-FIP element Ne. This method should account for any solar-cycle variation in density or temperature. Since we assumed photospheric abundances when deriving the DEM, the irradiance computed for the high-FIP element spectral line will be too high by factors of 3–4 if the corona has an enhanced composition of low-FIP elements. The ratio of the predicted-to-observed irradiance is therefore the FIP bias.

We used the CHIANTI database v.8 *(ref 37, 38)* for all the atomic data calculations. Uncertainties in these atomic data could also affect the FIP bias measurements. They are difficult to quantify, but we have assumed errors of 20% for each line. Our method has been specifically tested to assess the likely impact using EIS data *(33)*. In that experiment, a model DEM solution was used to calculate line intensities, which were then randomly perturbed within increasing error limits, and the dispersion in FIP bias measurements checked. The test showed that the FIP bias remains within 30% of the original value until the intensity errors approach 40–50%.

In our calculations, each FIP-bias measurement is computed from the best fit DEM solution from a sample of 100 MCMC runs. These individual runs adjust the DEM to produce the best fit to the observed irradiances, and we ensure that all of the observed irradiances are accurately reproduced for every data set by minimizing the $\chi^2$ (see Supplementary Figs. 5 and 6). With 14 observations, 4 smoothed independent temperature bins, and therefore 10 degrees of freedom, this approach results in reduced $\chi^2$ values of <1 due to the size of the irradiance uncertainties. This implies that the combination of (random) intensity calibration and assumed atomic data errors is not technically capturing the real nature and size of the uncertainties (which are expected to be systematic). It is difficult, however, to justify reducing them on purely technical grounds.

**Uncertainties and comparisons with previous work**

In order to determine the uncertainty in the measurements, we conducted an experiment where we re-computed the FIP bias in 1000 separate simulations for the data set used in Supplementary Fig. 5 i.e., we calculated FIP bias values for 1000 independently computed DEMs, each of which is the best-fit solution to 100 MCMC runs. From these computations, we determined the variation in the FIP bias (standard deviation of the distribution) resulting from variations in the best-fit DEM solutions. We show the results in Supplementary Fig. 7. The result is that the FIP-bias measurements have an uncertainty of ~ 0.3. We show this error bar on the linear scale of Fig. 2, and the logarithm reduced error bar on the log scale of Fig. 3.

There is some uncertainty in the values to adopt for the photospheric abundances. The values we used *(34)* are 7.53 for Mg, 7.51 for Si, 7.45 for Fe, and 7.84 for Ne. Note that adopting a different abundance for Ne will shift our FIP-bias plot up or down, but that the relative magnitude of the increase from minimum to maximum will be unchanged. The correlation with the F10.7 cm radio flux will also remain.

It is likely that stellar coronal abundance measurements, made using similar emission measure analysis techniques to ours, will also have uncertainties of a similar or even larger magnitude to our method (~ 0.3). The stellar coronal values we adopt (Fig. 3 in the main text constructed from multiple sources *(13,15,39,40,41,42,43,44,45)* are averages from the high-FIP elements C, N, O, and Ne. The spread in the original values for each element separately is also comparable to our experimentally determined error (~ 0.3) *(ref 15)*. Further, stellar photospheric abundance measurements do not always exist and are not always well determined. At least for the M stars in the sample, solar photospheric abundances were used, and the determinations for some of the others have large error bars *(15)*.

As mentioned in the introduction, there is also evidence of a cyclic dependence in the Ne/O ratio in the slow wind *(9)* and coronal streamers *(10)*, which show an increase at solar minimum. Conversely, older soft X-ray spectra show an increase with increasing active region temperature, which would suggest the opposite cyclic trend *(46)*. These and other measurements *(20, 47)* have demonstrated that variations in high-FIP element abundances exist, suggesting that some high-FIP elements may also be fractionated to some degree. Oxygen for example may behave more like a low-FIP element relative to Ne in coronal streamers *(10)*. The observations also recall the question of whether low-FIP elements are enhanced, or whether high-FIP elements are depleted. The behavior of the Ne/O ratio with active region temperature *(46)* may be explained by a depletion of Ne. If this occurred near solar maximum, it could produce a similar trend to what we observe.

If the variability of the Ne abundance is larger than for other high-FIP elements, this may also imply a larger cycle variation. Support for that possibility also comes from our results in comparison to the only other full-Sun EUV spectroscopic observation taken during the period of our observations *(33)*. Our measurement is higher (factor of ~ 2) than the average value from that study, suggesting that the variation we observe during the solar cycle may indeed be larger than detectable using other high-FIP elements. We caution that these results are not directly comparable: different elements (Ne vs. S) and data handling techniques (Sun-as-a-star vs. spatially resolved full-Sun maps) were used. In particular, S lies close to the boundary between low- and high-FIP elements and sometimes shows contradictory behavior. Spatially resolved maps also have more accurate measurements where the signals are strong, e.g., in active regions, leading to lower average values. While this implies that the cyclic dependence we are showing may be close to an upper limit, it also suggests that since Ne lines are the most sensitive, they might be the best observing tools for detecting any similar dependence in X-ray observations of solar-like stars.

The key issue for this work is determining the dominant behavior when the Sun is observed as a star. Analysis of full-Sun spectra *(48)* taken by several rocket flights between July 1963 and March 1969, covering the rise from the previous minimum through to around the maximum of solar cycle 20, does not show conclusive evidence either for or against a variation in the Ne/O ratio. It could be that high-FIP element fractionation is difficult to observe in full-disk integrated spectra. A study looking at Ne specifically (compared to Mg), using the SOHO CDS data from 1998 to 2010, states that little variation was found *(49)*. This study, however, did not directly compare the low- to high-FIP element ratio to the solar F10.7 cm radio flux, so it would be interesting to revisit these data in the future. We remind the reader that Ne and O are both high-FIP elements, so a lack of detection of fractionation in full-disk integrated spectra for those elements does not contradict our positive detection of fractionation between low- and high-FIP elements in our Sun-as-a-star observations.

**Data availability**

The data that support the findings of this study are available from the corresponding author upon request.

**Acknowledgements**

We thank Brian Wood for helpful comments on the paper. The work of D.H.B. was performed under contract with the Naval Research Laboratory and both D.H.B. and H.P.W. were funded by the NASA Hinode program. D.B. and L.v.D.-G. are funded under STFC consolidated grant number ST/N000722/1. L.v.D.-G. acknowledges the Hungarian Research Grant OTKA K-109276. The data are courtesy of the NASA/SDO and the AIA, EVE, and HMI science teams. CHIANTI is a collaborative project involving George Mason University, the University of Michigan (USA) and the University of Cambridge (UK).

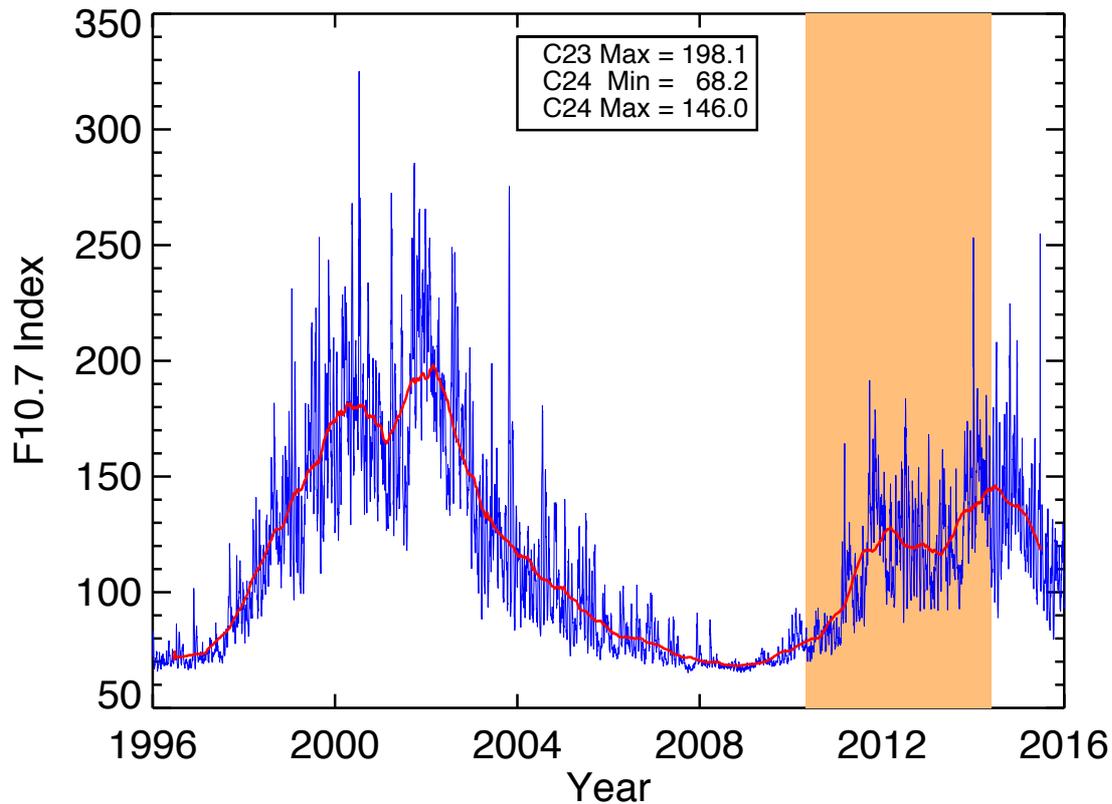

**Figure 1. Variation of the F10.7 cm radio flux between 1996 and 2016.** The F10.7 cm composite daily averaged solar radio flux index for the last 20 years (1996–2016). We show the evolution of the previous solar cycle 23 and current cycle 24 in blue. We also show the period of the EVE (Extreme-ultraviolet Variability Experiment) observations studied in this paper with a transparent orange overlay. The red line shows a boxcar smoothed yearly average fit to the two cycles, and the legend gives the F10.7 cm flux at the peak of cycle 23, the minimum between cycles, and the peak of cycle 24. The Figure also shows that the period of the EVE observations does capture the rise in solar activity from 2010 to 2014, even if the full solar cycle is unfortunately not sampled.

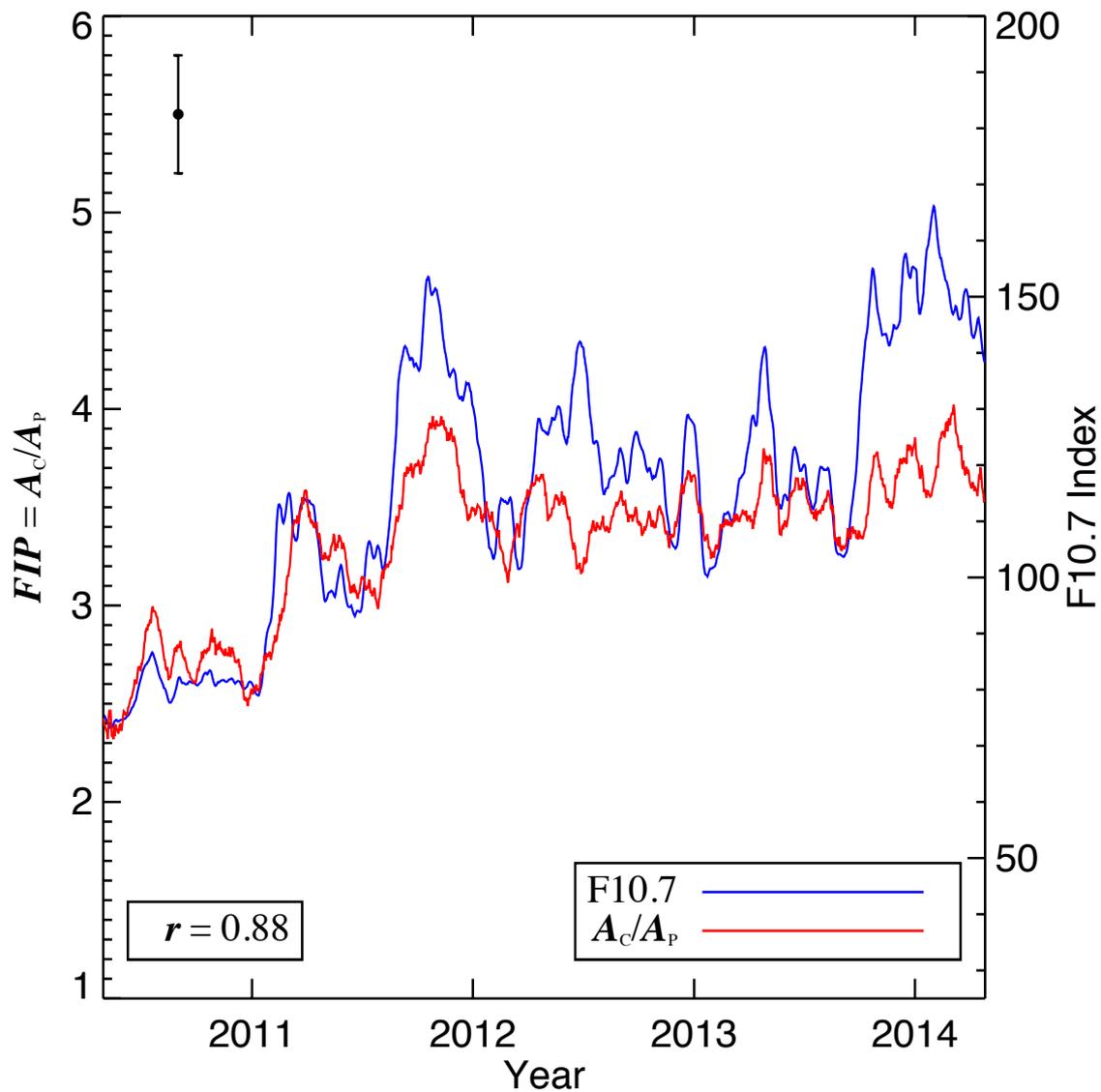

**Figure 2. Correlation between the coronal to photospheric composition ratio and the F10.7 cm radio flux.** The evolution of the 27-day Carrington rotation running average ratio of the coronal (Ac) to photospheric (Ap) composition in the Sun's corona between April 2010 and May 2014 (red) and the evolution of the 27-day Carrington rotation running average F10.7 cm radio flux (blue) during the same period. We show the correlation coefficient in the legend, and the uncertainty in the composition measurements with an error bar. Note that Ac/Ap is measured using the high first ionization potential (FIP) element Ne only. The increase in Ac/Ap appears most obvious when activity picks up after the end of the solar minimum (during 2010–11) and appears to level off thereafter while still closely tracking the F10.7 cm activity peaks. The error bar is the dispersion (standard deviation) in Ac/Ap computed from variations in the best fits to the irradiance data from a large number of random trials (see "Methods" section for more details).

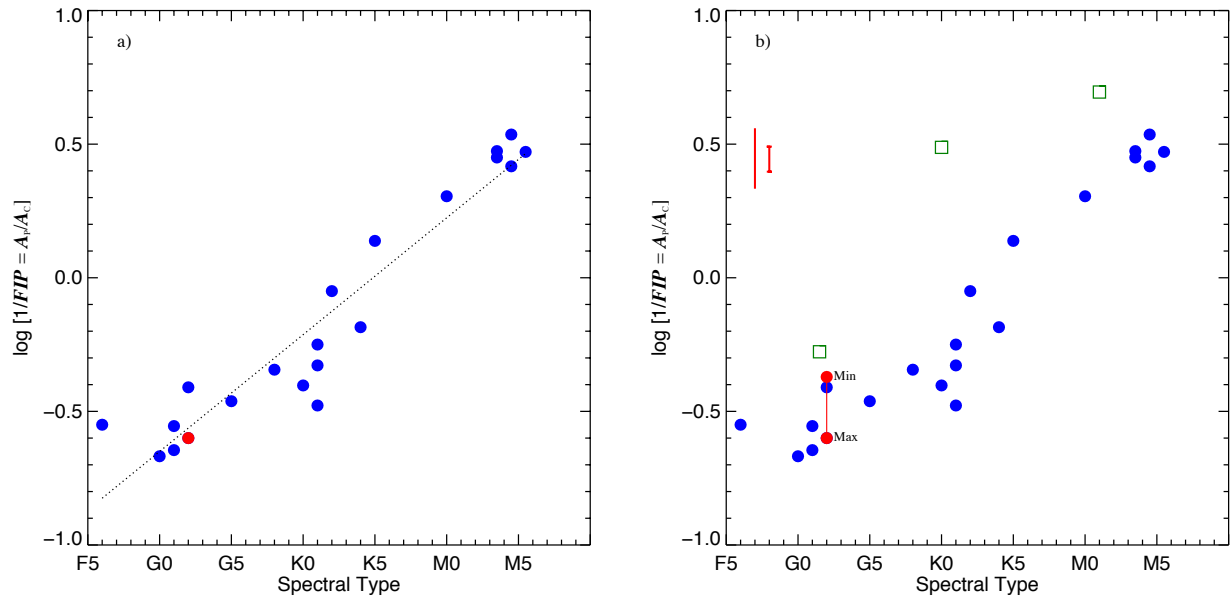

**Figure 3. Photospheric to coronal abundance ratios as a function of spectral type for a sample of stars. a** Logarithm of the ratio of photospheric to coronal abundance for a sample of stars as a function of spectral type. The data are adapted from Table 2 of Laming's review[16]. We show the Sun as a red dot. Note that the stellar values of Ac/Ap represent the average of the high first ionization potential (FIP) elements C, N, O, and Ne, whereas our measurements of Ac/Ap are based on Ne alone. The absolute values therefore may not be directly comparable. **b** We show the range of values for the Sun due to the solar-cycle variation. The solid red line shows the range of movement covered by the EVE (Extreme-ultraviolet Variability Experiment) measurements. We also show the same range of cyclic movement plotted as a solid bar in the top left for comparison with the stellar measurements. The smaller error bar is the uncertainty in our composition measurements. This is the dispersion (standard deviation) in Ac/Ap computed from variations in the best fits to the irradiance data from a large number of random trials (see "Methods" section for more details). The green squares are three very active stars (AB Dor, AU Mic, and EK Dra) that have been excluded from similar plots in the past.

**Supplementary Information**

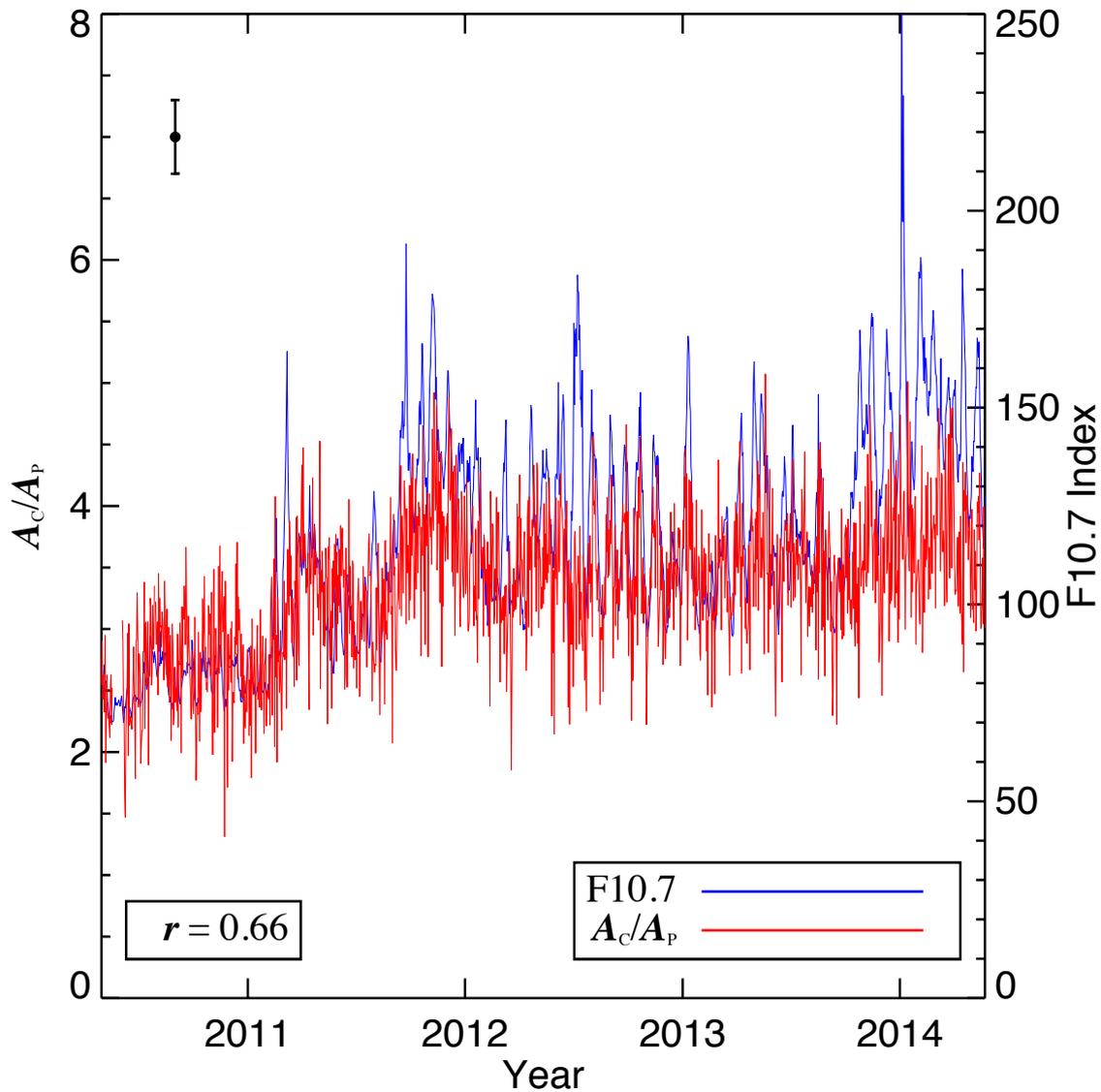

**Supplementary Figure 1: Daily running averaged composition and F10.7 cm data.** The evolution of the daily running averaged ratio of the coronal to photospheric composition in the Sun's corona between April 2010 and May 2014 (red) and the evolution of the daily running averaged F10.7 cm radio flux during the same period. We show the correlation coefficient in the legend. We also show the uncertainty in the composition measurements with an error bar. This is the dispersion (standard deviation) in $A_C/A_P$ computed from variations in the best fits to the irradiance data from a large number of random trials (see Methods for more details). The slightly lower correlation compared to Figure 2 may be because detailed daily variations due to activity, such as flares, are smoothed out in the Carrington 27-day running average data.

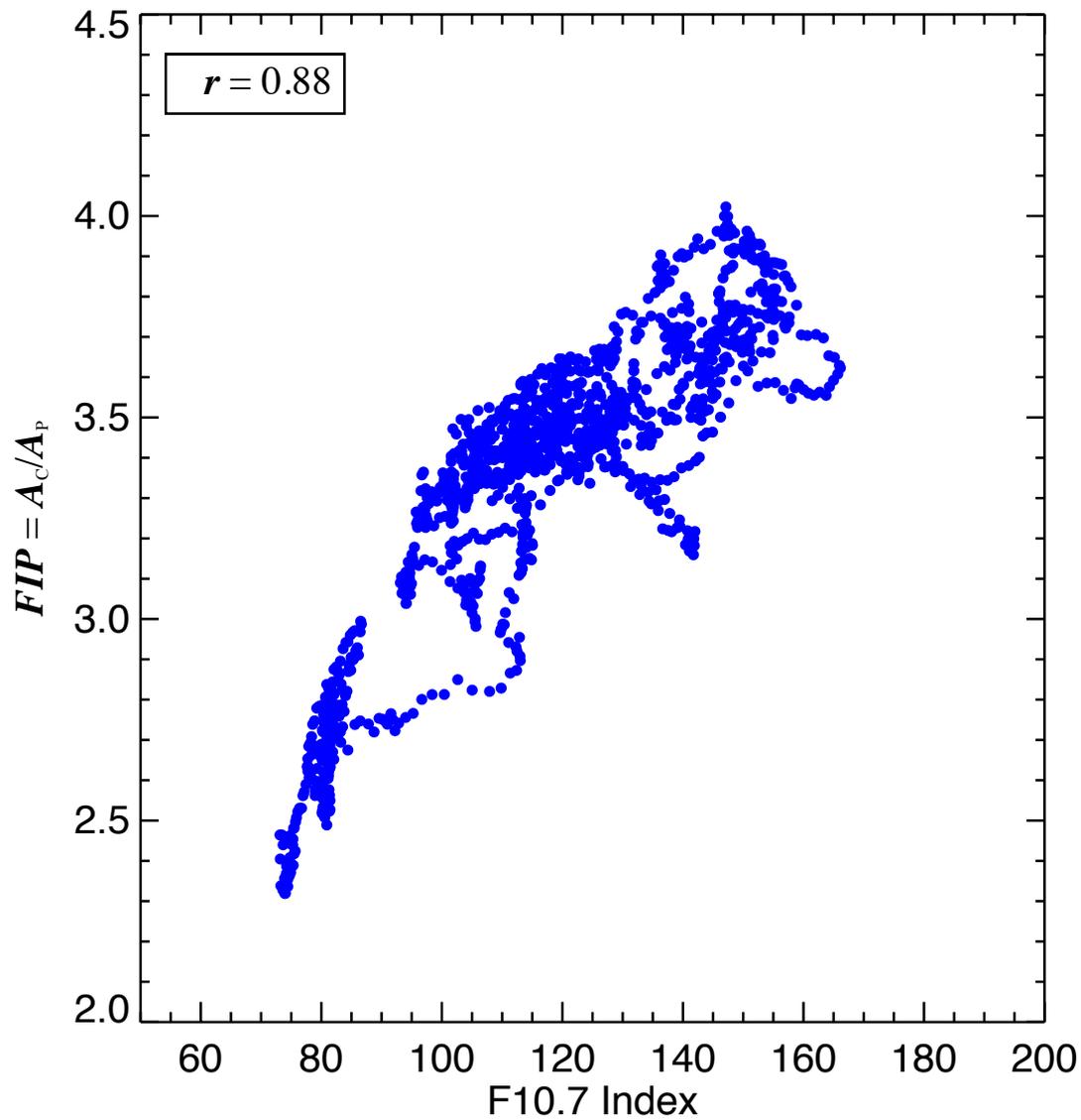

**Supplementary Figure 2: Correlation plot of the composition and F10.7 cm data.** Correlation between the 27-day Carrington rotation running averaged $A_C/A_P$ ratio and the F10.7 cm radio flux for April 2010 to May 2014. We show the correlation coefficient in the legend.

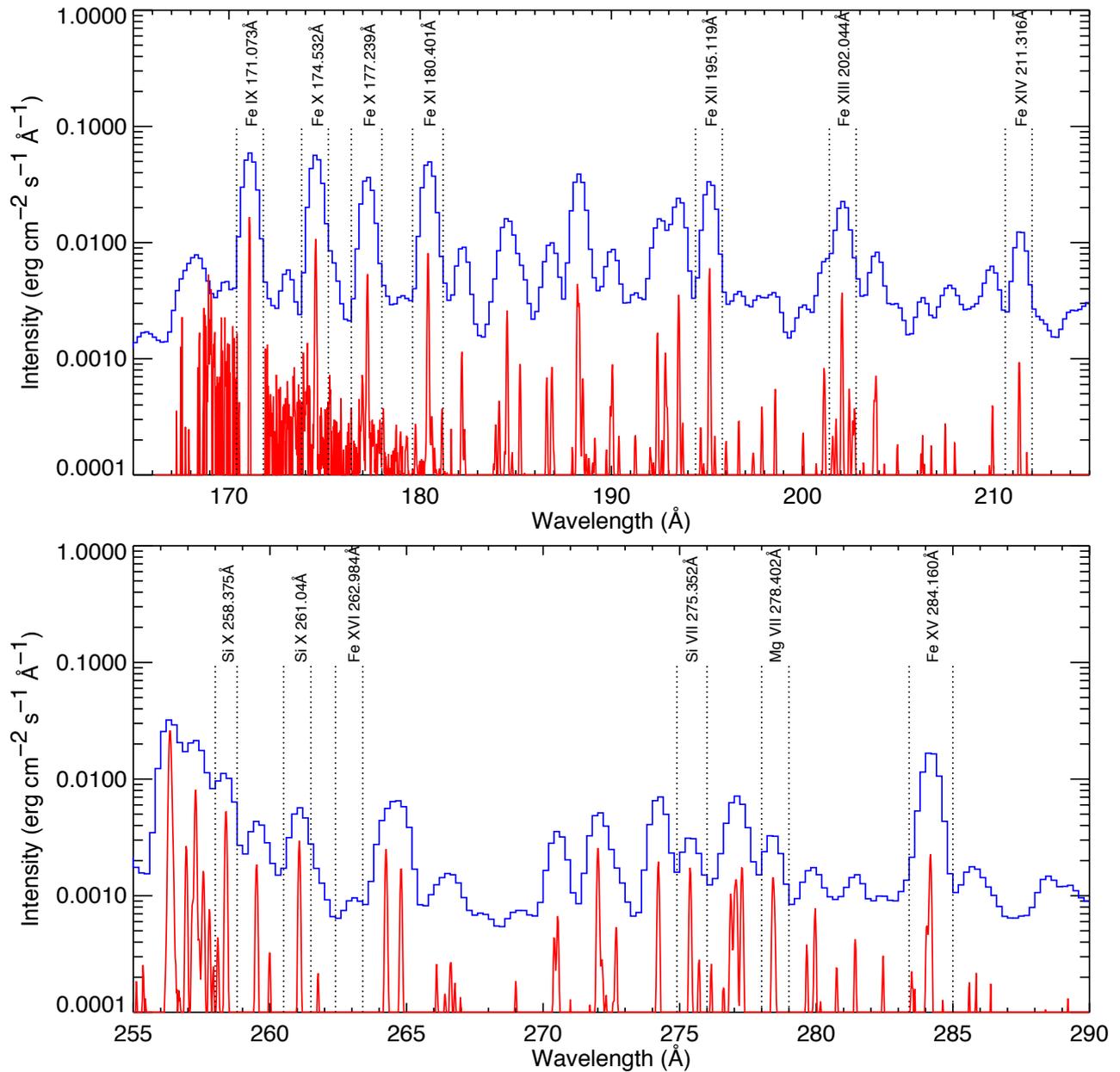

**Supplementary Figure 3: Example of the EVE spectra.** Example daily averaged EVE (Extreme ultraviolet Variability Experiment) spectra taken on 13th May 2010 (blue histogram). We have labelled the spectral lines used in the DEM (Differential Emission Measure) analysis, and the vertical dotted lines show the wavelength limits that were used to extract the intensities. We show a much higher spectral resolution quiet Sun EIS (Extreme-ultraviolet Imaging Spectrometer) spectrum for comparison (red line). The EIS spectrum was taken on 30th January 2007 using a much smaller field-of-view of 128 by 128 arcseconds.

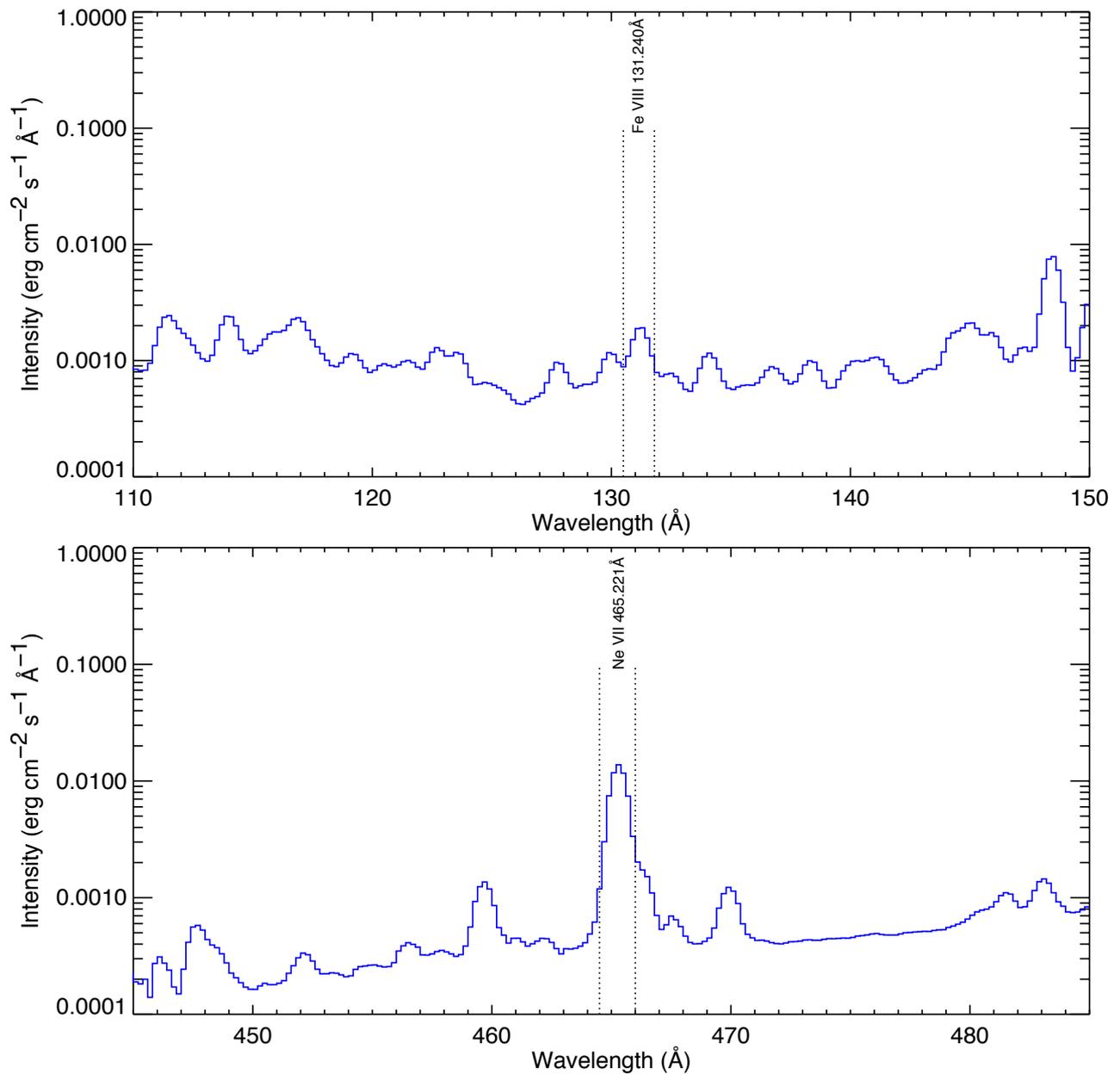

**Supplementary Figure 4: Example of the EVE spectra.** An example of the daily averaged EVE (Extreme ultraviolet Variability Experiment) spectra taken on 13th May 2010 (blue histogram) showing the wavelength intervals of the Fe VIII 131.24 Å line that was included to better constrain the lower temperatures, and the Ne VII 465.221 Å line that was used to measure the coronal to photospheric composition ratio.

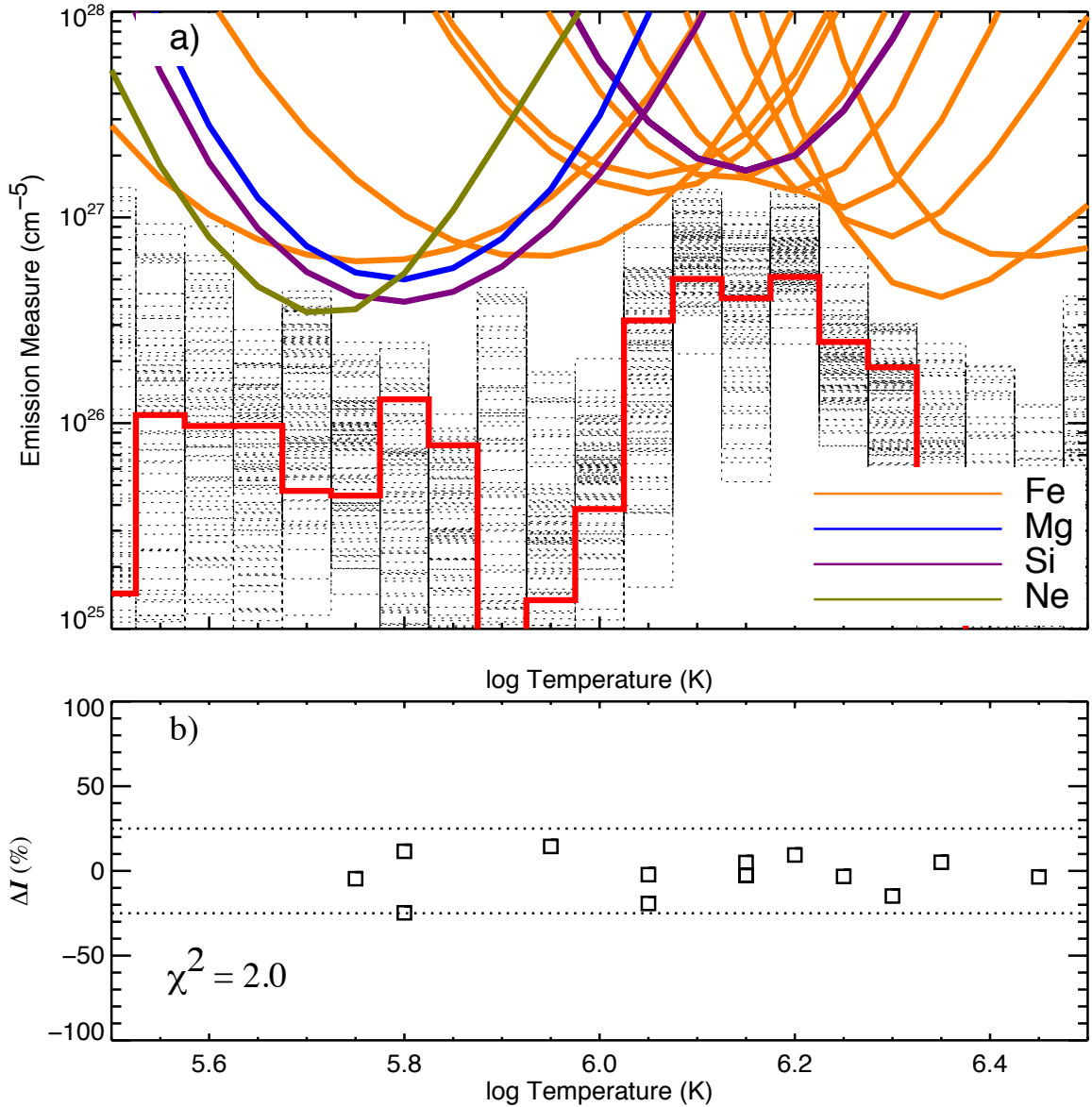

**Supplementary figure 5: Example of the coronal temperature disrtibution. a**, Example DEM (Differential Emission Measure) solution for the daily averaged spectrum taken on 13th May 2010. The dotted grey lines show the Monte Carlo simulations and the red line shows the best-fit solution. The colored lines are emission measure loci curves for all the spectral lines and show the upper limit constraints on the DEM. The Ne VII curve is the model solution. **b**, Differences between the observed and DEM calculated intensities expressed as a percentage of the observed intensity. The differences are all less than 25% and the resultant chi squared value for the fit is low (see legend).

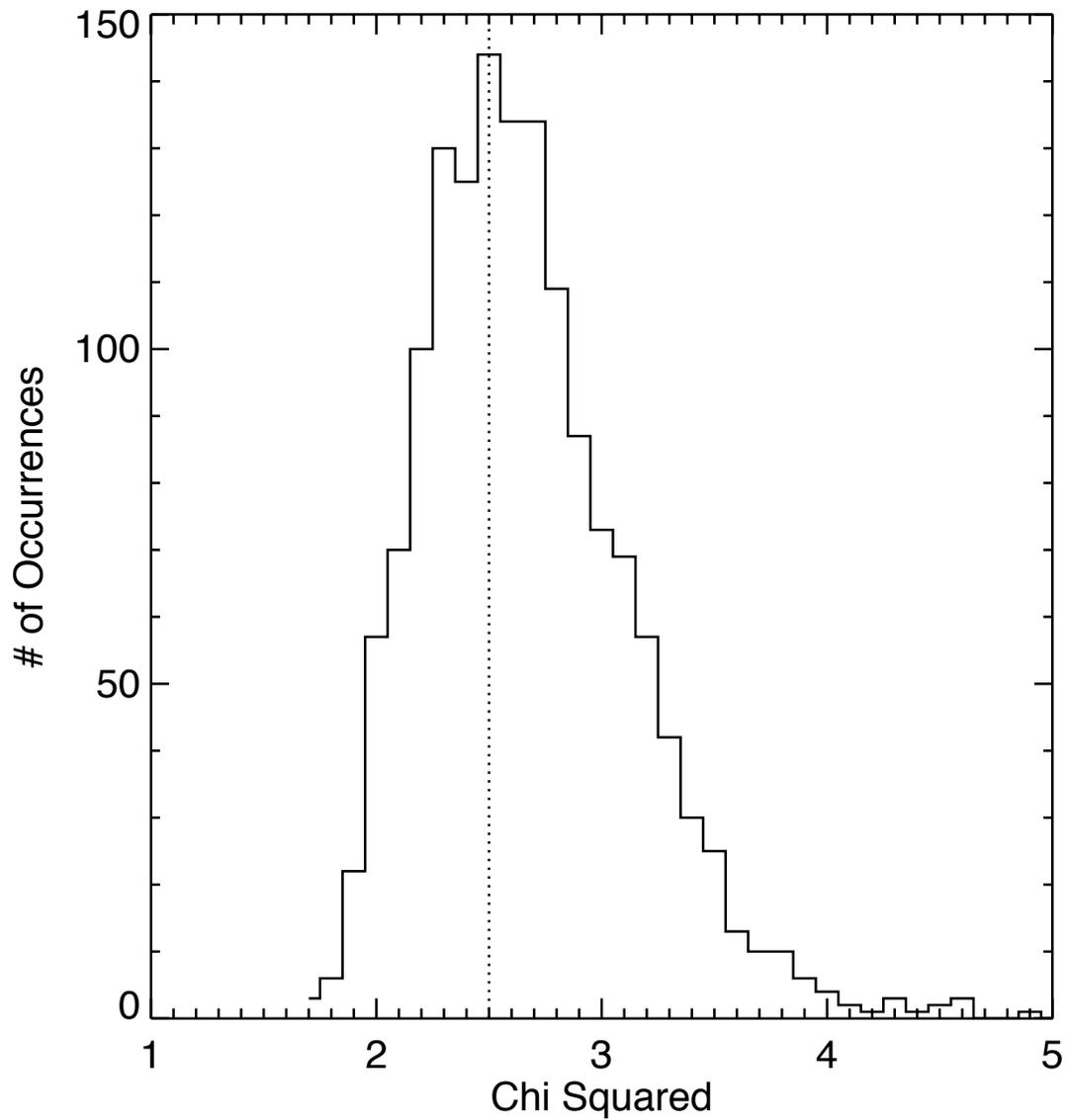

**Supplementary Figure 6: Chi squared values for the complete dataset.** Histogram of chi squared values for each best-fit DEM (Differential Emission Measure) solution to all the individual daily averaged spectra in the complete 4-year dataset. The results cover the time period from April 2010 to May 2014.

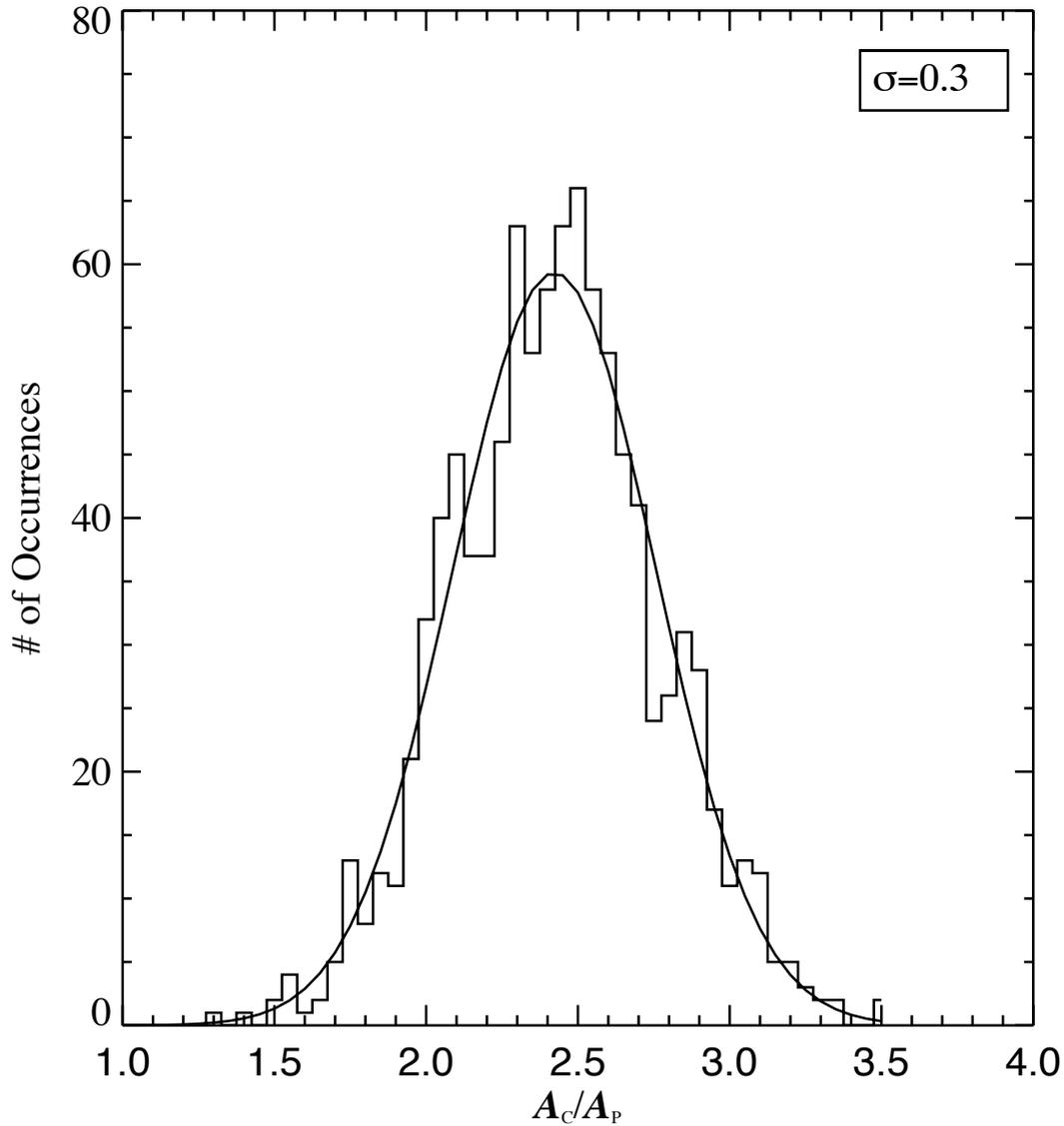

**Supplementary Figure 7: Example of the dispersion in the computed composition ratios.** Histogram of coronal to photospheric composition ratio for 1000 independent trial simulations for the 13th May 2010 dataset. Each of the 1000 measurements are computed from the best fit DEM (Differential Emission Measure) solution obtained from a sample of 100 MCMC (Markov-Chain Monte Carlo) simulations that adjust the DEM to fit the observed irradiances. We show the standard deviation, indicating the dispersion of the measurements, in the legend.